\documentclass[journal]{IEEEtran}
\ifCLASSINFOpdf
  \usepackage[pdftex]{graphicx}
\else
  \usepackage[dvips]{graphicx}
   \graphicspath{{../figures/}}
\fi
\usepackage{amsmath}
\pdfoutput=1
\begin{document}
%
\title{Impact of aperture, depth, and acoustic clutter on performance of Coherent Multi-Transducer Ultrasound imaging}
%
%
%

\author{Laura~Peralta,
        Alessandro~Ramalli,
        Michael~Reinwald,
        Robert J.~Eckersley,
        and~Joseph V.~Hajnal
\thanks{L. Peralta, M. Reinwald, R. J. Eckersley, and J.V. Hajnal are with the Department of Biomedical Engineering, School of Biomedical Engineering \& Imaging Sciences, King's College London, St Thomas' Hospital, Westminster Bridge Rd, SE1 7EH, London, UK e-mail: laura.peralta\_pereira@kcl.ac.uk }
\thanks{A. Ramalli is with the Department of Information Engineering, University of Florence, Florence, Italy.}
}

\maketitle

\begin{abstract}
Transducers with larger aperture size are desirable in ultrasound imaging to improve resolution and image quality. A coherent multi-transducer ultrasound imaging system (CoMTUS) enables an extended effective aperture through coherent combination of multiple transducers. In this study, the discontinuous  extended aperture created by CoMTUS and its performance for deep imaging and through layered-media are investigated by both simulations and experiments. Typical image quality metrics - resolution, contrast and contrast-to-noise ratio - are evaluated and compared with a standard single probe imaging system. Results suggest that, the image performance of CoMTUS depends on the relative spatial location of the arrays. The resulting effective aperture significantly improves resolution, while the separation between the arrays may degrade contrast. For a limited gap in the effective aperture (less than few centimetres), CoMTUS provides benefits to image quality compared to standard single probe imaging system. 
Overall, CoMTUS shows higher sensitivity and reduced loss of resolution with imaging depth. In general, CoMTUS imaging performance was unaffected when imaging through a layered-medium with different speed of sound values and resolution improved up to 80\% at large imaging depths.
\end{abstract}

\begin{IEEEkeywords}
multi-probe, large aperture, ultrasound imaging, acoustic clutter
\end{IEEEkeywords}

%
\IEEEpeerreviewmaketitle

\section{Introduction}
%
%
%
%
\IEEEPARstart{L}{imited} resolution and a restricted field of view (FoV) are two of the main challenges in ultrasound (US) imaging that, in principle, are fundamentally limited by the extent of the transmitting and receiving apertures. The image lateral resolution of ultrasonography is diffraction limited and mostly depends on the wavelength of the transmission wave, the imaging depth and the aperture size (measured by F-number) \cite{cobbold2006foundations}. However, since high frequency acoustic waves are easily absorbed and attenuated, the use of low frequencies is the only choice when the imaging region is located in deep areas. Consequently, a large aperture size is desired to decrease the F-number, improve resolution and FoV. 

Nevertheless, it is not clear that a large aperture will overcome the expected resolution loss through depth in the presence of tissue heterogeneity \cite{harris1991ultimate}. Inhomogeneities and tissue layers with different acoustic properties cause phase errors restricting the improvements provided by large arrays \cite{moshfeghi1988vivo}. Transabdominal US imaging is particularly difficult in obese patients because of the increased imaging depth with longer attenuation paths and the presence of clutter associated with boundaries between regions with dissimilar properties \cite{tsai2015obesity,klysik2014challenges}. Difficult abdominal imaging tasks require an array design that provides sufficient penetration for the chosen target depth, adequate resolution for the diagnostic task and tolerance to acoustic clutter effects that degrade the backscattered echo signals. The need for the development of large arrays for high resolution abdominal diagnostic imaging has been already suggested by previous studies, where a large array was synthesized through a synthetic aperture method by combining the received data from multiple positions of the same probe \cite{bottenus2016feasibility,bottenus2018evaluation}. However, in such a method the final coherent aperture is limited by the long acquisition times and the image quality is very sensitive to calibration errors and noise in the tracking system.

Despite a large aperture being desirable, the practical aperture size is often limited by the complexity and cost of scanners and probes based on large channel count arrays and the low flexibility for some applications. Using collections of conventional existing arrays to extend the aperture may be a way to approach this problem. Indeed, the use of multiple conventionally sized arrays to create a large aperture instead of using a single big array may be more flexible for diverse applications, e.g. for intercostal imaging, where the acoustic windows are narrow. In practice, multiple probes can be manipulated using a single and potentially adjustable holder that allows the operator to hold multiple probes with only one hand while keeping them directed to the same region of interest \cite{zimmer2018}. In that study incoherent image compounding was used, achieving benefits from view diversity but without directly achieving the gains in resolution and sensitivity that could be realised if the multiple probes were fully combined into a one large effective aperture. To achieve this requires coherent multi-transducer ultrasound (CoMTUS) imaging, which coherently combines all radio frequency (RF) data received by multiple synchronized transducers that take turns to transmit plane waves (PWs) into a common FoV \cite{peralta2018ius,peralta2019}. Hence, CoMTUS improves ultrasound imaging performance, yielding images that in general present an extended FoV and higher resolution along with a high acquisition frame rate. Previously coherent combination required precise telemetry to achieve the subwavelength localization accuracy required to combine information from multiple transducers poses \cite{bottenus2016feasibility,bottenus2018evaluation}, but \cite{peralta2019} demonstrated this can be achieved without the use of any external tracking device by optimizing the beamforming parameters. These optimal beamforming parameters, which include the transducer locations and the average speed of sound (SOS) in the medium, are deduced by maximizing the coherence of the received RF data resulting from different targeted scatterers in the medium by cross-correlation. CoMTUS provides the first step toward large aperture imaging using independently placed transducers to form discontinuous extended apertures with more flexibility than a conventional large probe may have.

One key feature of CoMTUS is the resulting discontinuous aperture. Preliminary studies showed that, as a consequence of the discontinuous aperture there is a trade-off between resolution and contrast \cite{peralta2019SPIE, peralta2019iusaberration}. However, how much the discontinuities dictated by the spatial separation between the multiple transducers may affect the global performance of the method is still unclear. Moreover, so far, CoMTUS has been validated only in a constant SOS media and little is known about its behavior in the case of different and spatially varying speeds of sound. Notwithstanding, since the average SOS is a beamforming parameter that is optimized along with the location of the probes, an improvement in beamformation with tolerance for some path dependent speed might be expected. The purpose of this study is to further investigate the feasibility of CoMTUS as different aspects of the geometry and material properties are varied to better assess its potential benefits and describe its implications for imaging.   


\section{\label{sec:2} Theory. Coherent multi-transducer ultrasound imaging}

A CoMTUS system consists of $N$ synchronized US arrays that take turns to transmit PWs into a common FoV. Only one individual array transmits at each time while all $N$ arrays simultaneously receive. The subwavelength localization accuracy required to coherently combine RF data from multiple probes is achieved without the use of any external tracking device.  
The method is briefly summarized in this section, but more details can be found in \cite{peralta2018ius,peralta2019,peralta2019iusextension}. 

\begin{figure}[htb!]
\centering
\includegraphics[width=\linewidth]{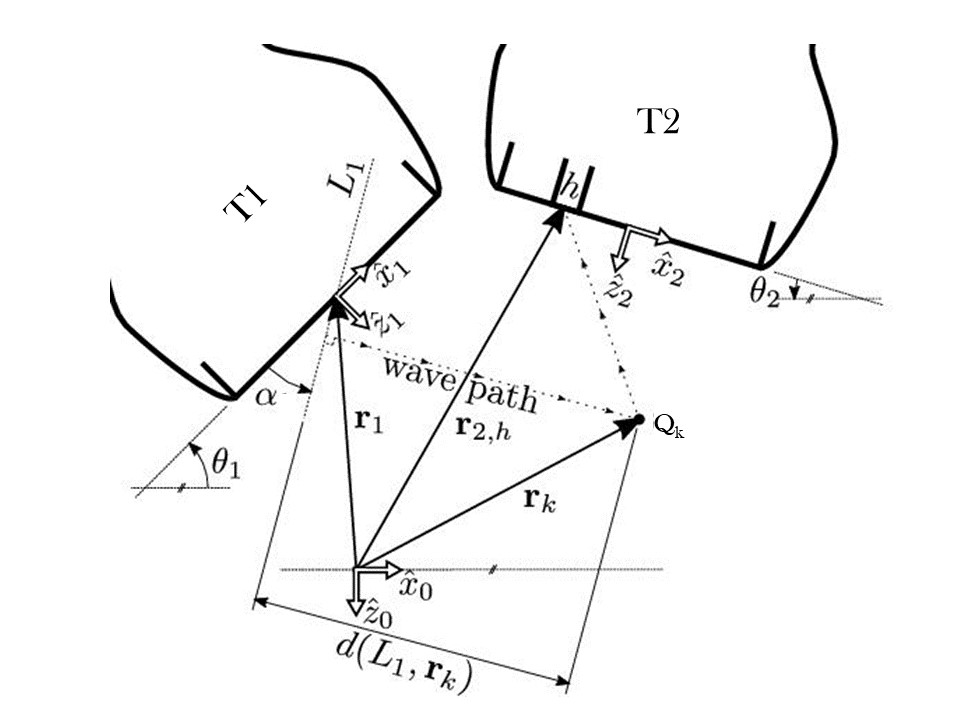}
\vspace{-0.5cm}
\caption{CoMTUS beamforming scheme. Global and local coordinates of arrays are shown. In this example, array T1 transmits a PW at angle $\alpha$ and array T2 receives the resulting echoes backscattered from the scatter $Q_k$.}
\label{fig:coordinate system}
\end{figure}

In this study, we consider a CoMTUS system formed by $N=2$ identical linear arrays that lie in the same elevational plane ($y=0$) and are angled relative to each other so as to share part of the FoV, as shown in Fig. \ref{fig:coordinate system}. In such a configuration, the following nomenclature is used.  For a sequence in which array $i$ transmits and array $j$ receives, the RF data received on channel~\(h\) of array $j$ at time~\(t\) is named \(T_iR_j(h,t)\). The resulting image and all transducer coordinates are defined in a global coordinate system arbitrarily located in space $(\hat{x}_{0},\hat{z}_{0})$. For a given linear array $i$, a local coordinate system~\((\hat{x}_{i},\hat{z}_{i})\) is defined at the center of the array surface with the~\(\hat{z}\) direction orthogonal to the transducer surface and directed away from array $i$.
The position and orientation of array $i$ are then characterized in the global coordinate system with 3 parameters that define a translation vector $\mathbf{r}_{i}$ and a rotation angle $\theta_{i}$ \cite{fitzgibbon2003robust}.

Standard delay and sum beamforming for PW imaging \cite{Montaldo2009CoherentElastography} can be extended to the present multi-transducer set-up taking into account the full path length between the transmit array and the receive elements. Considering that array $i$ transmits a PW at certain angle $\alpha$, the image point to be beamformed located at $Q_k$ and described by the vector $\mathbf{r}_{k}$ can be computed from the echoes received at transducer $j$ as:
\begin{equation}
\begin{split}
s_{i,j}(Q_k;\alpha) = \sum_{h=1}^{H} T_iR_j\left(h,Q_k;\alpha\right) = \\  
\sum_{h=1}^{H} T_iR_j\left(h,\frac{D_{i,j,h}(Q_k;\alpha)}{c}\right)
\end{split}
\label{eq:beamforming}
\end{equation}
where $H$ is the total number of elements in the array and $c$ is the average SOS of the propagation medium. The total distance $\mathrm{D}_{i,j,h}(Q_k)$ between the transmit array $i$, the imaging point $Q_k$ and the receive element $h$ of array $j$ is defined by, 
\begin{equation}
\mathrm{D}_{i,j,h}(Q_k;\alpha) = d(L_i(\alpha),\mathbf{r}_k) + \lVert \mathbf{r}_{j,h}-\mathbf{r}_k  \rVert
\end{equation}
where $d(\mathbf{r}_k,L_i(\alpha))$ is the distance from $\mathbf{r}_k$ to the line $L_i(\alpha)$ that defines the transmitted PW, and $\lVert \mathbf{r}_{j,h}-\mathbf{r}_k \rVert$ is the Euclidean distance between $\mathbf{r}_k$ and the receive element $h$ of array $j$. These distances are represented in Fig. \ref{fig:coordinate system}.

Finally, the total beamformed image $S(Q_k;\alpha)$ can be obtained by coherently adding the individually beamformed images acquired in a sequence in which both probes transmit: 
\begin{equation}
\begin{split}
S(Q_k;\alpha) & = 
 s_{1,1}(Q_k;\alpha) + s_{1,2}(Q_k;\alpha)  \\
 & + s_{2,1}(Q_k;\alpha) + s_{2,2}(Q_k;\alpha)
\end{split}
\label{eq:finalImage}
\end{equation}
In the same way, several PWs transmitted at different angles, $a=1,\dots,A$ may be coherently combined to generate an image,
\begin{equation}
S(Q_k;A) = \sum_{i=1}^N \sum_{j=1}^N \sum_{a=1}^A s_{i,j}(Q_k;\alpha_a^{i})
\label{eq:finalImageAngle}
\end{equation}

The optimum beamforming parameters that determine equation (\ref{eq:beamforming}) ($\mathcal{P} = \{c,\theta_1,\mathbf{r}_1, \theta_2,\mathbf{r}_2\}$), not a priori accurately known, are calculated by maximizing the cross-correlation of backscattered signals from common point-like targets acquired by individual receive elements, by using gradient-based optimization methods,
\begin{equation}
\bar{\mathcal{P}} = \arg \max_\mathcal{P} \chi(\mathbf{\mathcal{P}})
\label{eq:optimization}
\end{equation}
where the cost function $\chi(\mathcal{P})$ is defined as follows,   
\begin{equation}
\begin{split}
\chi(\mathcal{P}) &= \\
&\sum\limits_{k}^{K} \sum\limits_{h}^{H} \{ \text{NCC}(T_1R_1(h,Q_k;\mathcal{P}),T_2R_1(h,Q_k;\mathcal{P}))W_{1,1}W_{2,1}
 \\ 
 &+  \text{NCC}(T_1R_2(h,Q_k;\mathcal{P}),T_2R_2(h,Q_k;\mathcal{P}))W_{1,2}W_{2,2} \}
\end{split}
\label{eq:cost}
\end{equation}
NCC is the normalized cross-correlation and $W_{i,j}$ is a weighting factor defined as,
\begin{equation}
W_{i,j}(\mathbf{\mathcal{P}}) = \frac{1}{2} + \frac{1}{2H}\sum_{h_b \neq h}^{H} \text{NCC}(T_iR_j(h;\mathcal{P}),  T_iR_j(h_b;\mathcal{P})) 
\label{eq:weights}
\end{equation} 

A minimum of 2 points are needed to solve this trilateration problem. Although since the approach relies on a measure of coherence, which may well be more tolerant, previous results suggest that the
method may work when there are identifiable prominent local features. On the other hand, isolated point scatterers can be artificially generated by other techniques, for instance by inclusion of microbubble contrast agents \cite{christensen2014vivo}. Preliminary results show the potential of microbubble to optimize the beamforming parameters in CoMTUS, which could be implemented in vivo wherever the FoV contains vasculature \cite{christensen2019coherent}.


\section{\label{sec:3}Materials and Methods}
\subsection{\label{sec:3.1}Simulations}
The k-Wave Matlab toolbox was used to simulate the non-linear wave propagation through an inhomogeneous dispersive medium \cite{treeby2012modeling,treeby2010k}. A CoMTUS system formed by two identical linear arrays, similar to the ones experimentally available, was simulated as follows. Each of the arrays had a central frequency of 3 MHz and 144 active elements in both transmit and receive, with element pitch of 240 $\mu$m and kerf of 40 $\mu$m. For PWs the modelled transducer had an axial focus of infinity with all 144 elements firing simultaneously. 
A simulation was performed for each transmit event, i.e. each PW at a certain angle. In total 7 transmit simulations per linear array were performed to produce a PW data set, which covers a total sector angle of 30$^{o}$ (from -15$^{o}$ to 15$^{o}$, 5$^{o}$ step). In the case of CoMTUS this results in 14 transmit events in total (7 PWs per array). This PW sequence was chosen to match in resolution a focused system with F-number 1.9, decimating the required number of angles by a factor of 6 to optimize the simulation time without affecting resolution \cite{denarie2013coherent}. The computational grid was set following convergence testing from a previous study \cite{Brown2019Detection}. The spatial grid was fixed at 40 $\mu$m (six grid points per wavelength) with a time step of 1.3 ns. 
The power law exponent to model frequency-dependent attenuation was set to 1.5 \cite{treeby2010a}. 
Received signals were downsampled at 30.8 MHz. Channel noise was introduced to the RF simulated data as white Gaussian noise. 
The signal to noise ratio (SNR) for all simulated data was set to 35dB at 50 mm imaging depth.

The US pulses were propagated through heterogeneous scattering media using tissue maps defined by the SOS, density, attenuation and nonlinearity (see Table \ref{tab:properties}). A medium defined only with the properties of general soft tissue was used as control case. 
To model the scattering properties observed in vivo, sub-resolution scatterers were added to the tissue maps. A total of 15 scatterers of 40 $\mu$m diameter, with random spatial position and amplitude (defined by a 5\% difference in SOS and density from the surrounding medium), were added per resolution cell, in order to fully develop speckle \cite{pinton2009heterogeneous}. Three point-like targets and an anechoic lesion were included in the media to allow the measurement of the basis metrics for comparing the imaging quality for different scenarios. A circular anechoic lesion of 12 mm diameter located at the center of the aperture of the common FoV, was modeled as a region without scatterers. The point-like targets were simulated as circles of 0.2 mm diameter with a 25\% difference in SOS and density with the surrounding tissue to generate appreciable reflection. The same realization of scatterers was superimposed on all maps and through the different simulations to keep the speckle pattern in the CoMTUS system, so any changes in the quality imaging metrics are due to changes in the overlying tissues, the imaging depth and the acoustical field. 

\begin{figure}
\centering
\includegraphics[scale=1]{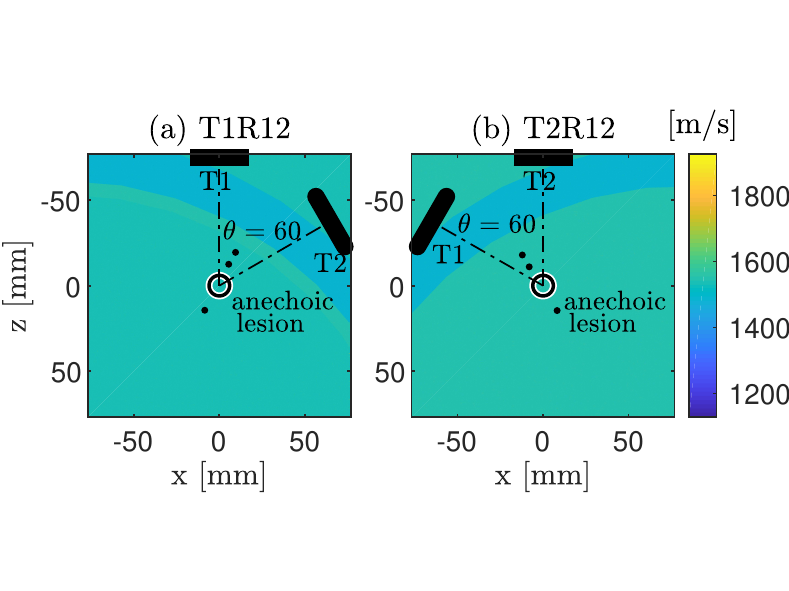}
\vspace{-1cm}
\caption{Example of the SOS map of a propagation medium with a muscle layer of 8 mm and a fat layer of 25 mm thickness. Locations of US probes, point-like targets and anechoic lesion are shown. (a) Medium expressed in the local coordinate system of the array T1 and used to simulate the RF data T1R12, i.e. when the array T1 transmits. (b) Medium expressed in the local coordinate system of the array T2 and used to simulate the RF data T2R12, i.e. when the array T2 transmits.
In this example, the angle between the probes that defines their position in space is 60$^{o}$ and the corresponding imaging depth 75 mm.}
\label{fig:medium}
\end{figure}
The k-Wave Matlab toolbox uses a Fourier co-location method to compute spatial derivatives and numerically solve the governing model equations, which requires discretisation of the simulation domain into an orthogonal grid. Consequently, continuously defined acoustic sources and media need to be sampled on this computational grid, introducing staircasing errors when sources do not exactly align with the simulation grid \cite{robertson2017accurate,wise2017staircase}. 
To minimize these staircasing errors, the transmit array was always aligned to the computational grid, i.e. simulations were performed in the local coordinate system of the transmit array. This implies that to simulate a sequence in which the array T2 transmits, the propagation medium, including the sub-resolution scatterers, was converted into the local coordinate system of probe T2 using the same transformation matrix that defines the relative position of both arrays in space.  
A sample tissue map with the transducers, targets and anechoic lesion locations, represented in both local coordinate systems, is shown in Fig. \ref{fig:medium}.

\subsubsection{CoMTUS discontinuous effective aperture}
\begin{figure}
\centering
\includegraphics[scale=1]{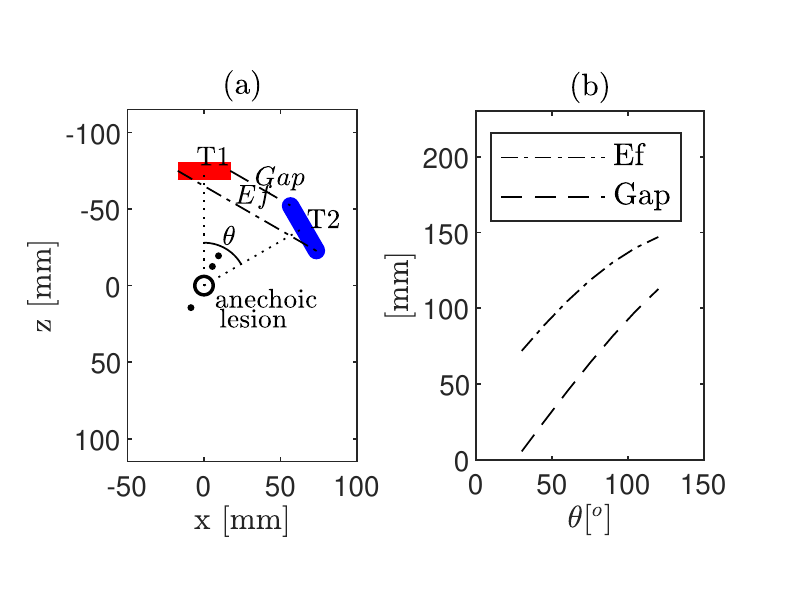}
\vspace{-0.5cm}
\caption{(a) Schematic representation of the spatial location of the two linear arrays, T1 (red) and T2 (blue), three point-like targets and the anechoic lesion in the local coordinate system of T1. Different geometrical parameters are shown: angle between probes ($\theta$) and the gap ($Gap$) in the resulting effective aperture ($Ef$). (b) Associations between $\theta$, $Gap$ and resulting $Ef$.}
\label{fig:geometry}
\end{figure}

Previous results showed that the discontinuous effective aperture obtained by CoMTUS determines the quality of the resulting image \cite{peralta2018ius,peralta2019SPIE}. To investigate the effects of the discontinuous aperture, determined by the relative location of the CoMTUS arrays in space, different CoMTUS systems with the arrays located at different spatial locations were modeled. Simulations were performed in the same control medium, where only soft tissue material was considered.
To modify the relative location of the probes while keeping the imaging depth (fixed at 75 mm), the angle between the arrays was changed. The array T1 was always positioned at the center of the $x$-axis of the simulation grid  while the array T2 was rotated around the center of the propagation medium. Then, different cases of CoMTUS with two arrays located at different angles ($30^{o}, 45^{o}, 60^{o}$ and $75^{o}$) were simulated. Fig. \ref{fig:geometry}(a) shows a schematic representation of the probes in space, where the different spatial parameters (angle between probes, $\theta$, and gap, $Gap$, in the resulting effective aperture, $Ef$) are labeled.  Note that, at larger angles, both the effective aperture of the system defined by both probes and the gap between them increase. The relationships between probe position, and the resulting effective aperture and gap are shown in Fig. \ref{fig:geometry}(b).

\subsubsection{CoMTUS image penetration}
\begin{figure}[h!]
\centering
\includegraphics[scale=1]{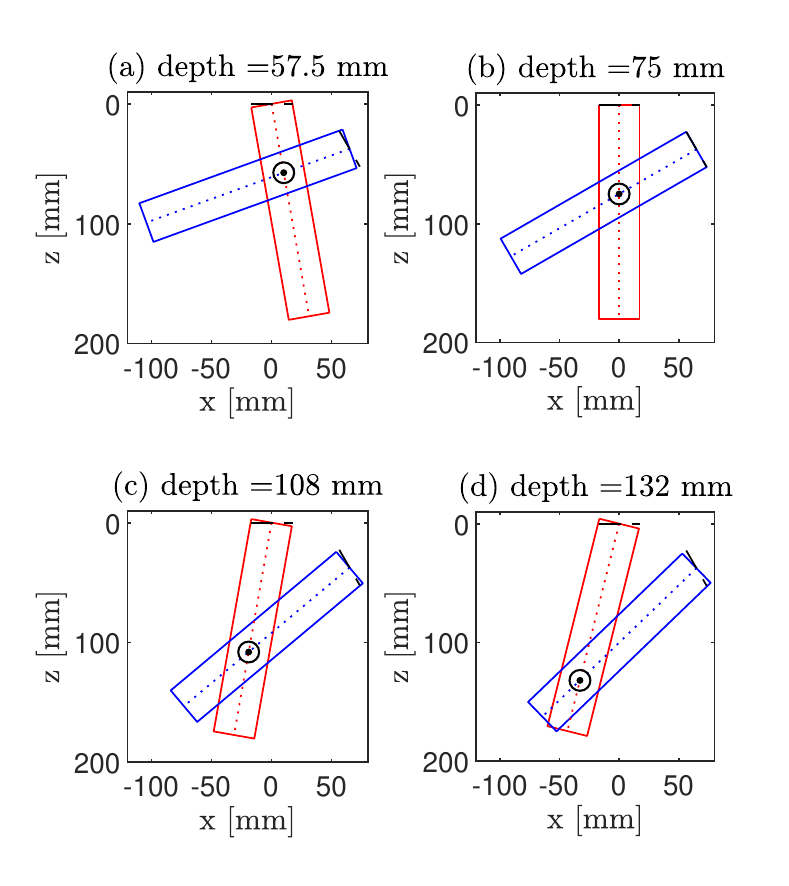}
\vspace{-0.5cm}
\caption{Schematic representation of the spatial location of the two linear arrays, T1 (red) and T2 (blue), and their FoV at different imaging depth. The circle indicates the center of the common FoV, which defines the imaging depth in CoMTUS. Four different imaging depths are shown: (a) 57.5 mm; (b) 75 mm; (c) 108 mm and; (d) 132 mm.}
\label{fig:depth}
\end{figure}

The image penetration of CoMTUS was investigated by changing the local orientation of the arrays and using the same control propagation medium (only soft tissue). For a given effective aperture (fixed gap $\sim$45.3 mm), each probe was rotated around its center the same angle but in the opposite direction. In that way, a certain given rotation, for example negative in T1 and positive in T2 will result in a deeper common FoV, and the opposite for the counter rotation. Fig. \ref{fig:depth} shows the imaging depth dependence on the transducer orientation (defined by the position of the common FoV of both arrays). Using this scheme, four different imaging depths were simulated: 57.5 mm, 75 mm, 108 mm, and 132 mm.

\subsubsection{CoMTUS through layered-media}
To investigate the effect of layers with different speed of sound in the medium, three different kinds of tissue were defined in the propagation media (general soft tissue, fat and muscle). The imaging depth was set to 75 mm with a configuration of the arrays in space that defines an effective aperture of 104.7 mm with 45.3 mm gap.  
The acoustic properties assigned to each tissue type were chosen from the literature \cite{goss1978comprehensive} and are listed in Table \ref{tab:properties}. A medium defined only with the soft tissue properties was used as control case. Then, clutter effects were analysed by using heterogenous media in which two layers with the acoustic properties of muscle and fat were introduced into the control case medium. In the different studied cases, the thickness of the muscle layer was set to 8 mm while fat ranged from 5 to 35 mm thickness. Fig. \ref{fig:medium} shows an example of the propagation medium with a muscle layer of 8 mm thickness and a fat layer of 25 mm.
\begin{table}[ht]
\caption{  Tissue map properties.}
\vspace{-0.5cm}
\label{tab:properties}
\begin{center}    
\begin{tabular}{|l|c|c|c|c|} 
\hline
\rule[-1ex]{0pt}{3.5ex}  Tissue& SOS & Density  & Attenuation & Nonlinearity  \\
\rule[-1ex]{0pt}{3.5ex} type & [m/s] & [kg/m$^3$]  & [dB/MHz/cm] & B/A  \\
\hline
\rule[-1ex]{0pt}{3.5ex} Soft tissue & 1540 & 1000  & 0.75  & 6 \\
\hline 
\rule[-1ex]{0pt}{3.5ex} Fat & 1478  & 950   & 0.63  & 10\\
\hline 
\rule[-1ex]{0pt}{3.5ex} Muscle & 1547 & 1050  & 0.15  & 7.4 \\
\hline 
\end{tabular}
\end{center}
\end{table}

Aberration parameters were estimated using a one-dimensional version of the reference waveform method \cite{liu1994correction}. In these measurements, a source located at the center of the shared FoV of both arrays that emits a spherical wave pulse that propagates through the different tissue layers and is received by the CoMTUS effective aperture (i.e. the 2 linear arrays) was simulated. Then, the arrival time fluctuations (ATF) across the receiving effective aperture caused by each layered medium were calculated by subtracting a linear fit from these calculated arrival times. Energy level fluctuations (ELF) in the data were calculated by summing the squared amplitudes of each waveform over a 1.5 ms window which isolated the main pulse, converting to decibel units, and subtracting the best linear fit from the resulting values. More details of these measurements can be found in \cite{mast1997simulation}. 

\subsection{\label{sec:3.2}In vitro Experiments}
A sequence similar to the one used in simulations was used to image a phantom. The imaging system consisted of two 256-channel Ultrasound Advanced Open Platform (ULA-OP 256) systems (MSD Lab, University of Florence, Italy) \cite{boni2016ula,boni2017architecture}. The systems were synchronized, i.e. with the same trigger and sampling times in both transmit and receive mode. Each ULAOP 256 system was used to drive an ultrasonic linear array made of 144 piezoelectric elements with a 6 dB bandwidth ranging from 2 MHz to 7.5 MHz (imaging transducer LA332, Esaote, Firenze, Italy). The two probes were mounted on xyz translation and rotation stage (Thorlabs, USA) and were carefully aligned in the same elevational plane (y = 0). For each probe in an alternating sequence, i.e. only one probe transmits at each time while both probes receive, 7 PWs, covering a total sector angle of 30$^{o}$ (from -15$^{o}$ to 15$^{o}$, 5$^{o}$ step), were transmitted at 3 MHz and pulse repetition frequency (PRF) of 1 kHz. RF data backscattered up to 135 mm deep were acquired at a sampling frequency of 19.5 MHz.  No apodization was applied either on transmission or reception.

A subset of the simulated results was experimentally validated in vitro. A phantom custom made as described in \cite{peralta2019}, with three point-like targets and an anaechoic region, was imaged with the imaging system and pulse sequences described above. The averaged SOS of the phantom was 1450 m/s. The phantom was immersed in a water tank to guarantee good acoustic coupling. To model a layered-medium with changes in SOS, a layer of paraffin wax of 20 mm thickness was placed between the probes and the phantom. The measured SOS of paraffin wax was 1300 m/s. The control experiment was performed first without the paraffin wax sample present. After the control scan, the paraffin wax sample was positioned over the phantom without movement of the phantom or tank. Then, the target was scanned as before. The paraffin wax sample was positioned to sit immediately over the phantom, coupled to the transducers by water. A final control scan was performed to verify registration of the phantom, tank and transducers, after the paraffin wax sample was scanned and removed. 

\subsection{Data processing}
The RF data, both simulated and experimentally acquired, were processed in different combinations to study image quality. 
The multi-transducer beamforming was performed as described in Section \ref{sec:2}, equations (\ref{eq:beamforming})-(\ref{eq:finalImage}) and the optimum beamforming parameters, calculated as described in Section \ref{sec:2} (equation (\ref{eq:optimization})), were used to generate CoMTUS images. 
For the simulated RF data, where the actual position of the arrays in space is known, an additional image, noted as 2-probes, was beamformed by assuming a SOS of 1540 m/s and using the spatial location of the array elements. Note that, in the experimental case this is not possible because the actual position of the arrays in space is not accurately known. Finally, the data corresponding to the sequence when the array T1 transmits and receives, i.e. $T_1R_1$, and noted here as 1-probe, was used as a baseline for array performance, providing a point of comparison to the current coherent PW compounding method \cite{Montaldo2009CoherentElastography} in both simulated and experimental scenarios. Note that, for all the cases except CoMTUS, an assumed value of the SOS was used to beamform the data (1540 m/s for simulated data and 1450 m/s for experimental data). 

In order to achieve a comparison between imaging modalities that is as fair as possible in terms of transmitted energy, the CoMTUS and the 2-probes images are obtained by compounding only 6 different PWs, while the 1 probe system images are generated compounding the total number of the transmit PWs, i.e. 7 PWs from $-15^{o}$ to $15^{o}$, in $5^{o}$ step. Specifically for the CoMTUS and 2-probes images, the results use compounded RF data when the array T1 transmits PW at zero and positive angles (0$^{o}$,5$^{o}$,10$^{o}$) and the array T2 transmits PW at zero and negative angles (0$^{o}$,-5$^{o}$,-10$^{o}$). An even number of transmissions was set because the CoMTUS optimization is based on pairs of transmissions, one per array (equation (\ref{eq:cost})). In addition, firing at opposite angles with the 2 arrays guarantees the CoMTUS performance since an overlap of the isonated regions is mandatory to determine the relative probe-to-probe position \cite{peralta2019}. 

For each resulting image, lateral resolution (LR), contrast and contrast-to-noise ratio (CNR) were measured to quantify the impact of both the aperture size and the clutter. LR was calculated from the point-spread-function (PSF) of the middle point target. An axial-lateral plane for 2-D PSF analysis was chosen by finding the location of the peak value in the elevation dimension from the envelope-detected data. Lateral and axial PSF profiles were taken from the center of the point target and aligned with the principal resolution directions. LR was then assessed by measuring the width of the PSF at the $-6$ dB level. 
The contrast and CNR were measured from the envelope-detected images \cite{smith1985frequency}. Contrast and CNR were calculated as: $\text{Contrast} = 20\log_{10}(\mu_i/\mu_o)$, and  $\text{CNR} = \mid \mu_i-\mu_o \mid / \sqrt{\mu_{i}^{2}+\mu_{o}^{2}}$, where $\mu_i$ and $\mu_o$ are the means of the signal inside and outside of the region, respectively, and at the same imaging depth. A circular region of 10 mm diameter was used for the these calculations. 


\section{\label{sec:4}Results}
\subsection{Simulation results}
\subsubsection{Control case: conventional aperture imaging}
\begin{figure*}
\centering
\includegraphics[width=1\linewidth]{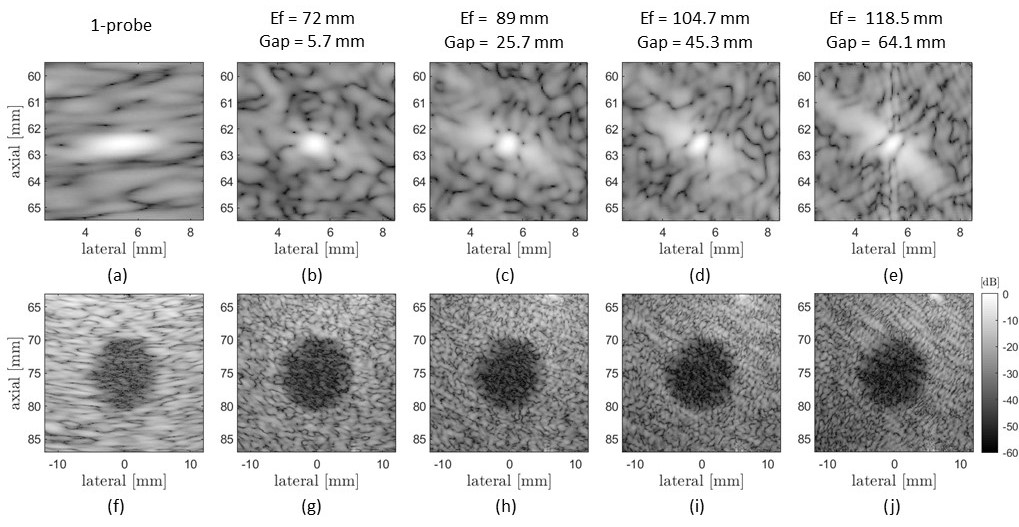}
\vspace{-0.5cm}
\caption{Simulated point targets and lesion images of the control medium obtained by 1-probe (a,f) and CoMTUS (b-e, g-j) with different effective aperture (Ef) and gap between the two arrays (Gap).}
\label{fig:CMTUSangle}
\end{figure*}
The conventional aperture image, corresponding to the sequence when the array T1 transmits and receives (1-probe) provides the base line for imaging quality through the different scenarios. 
Fig. \ref{fig:CMTUSangle}(a,f) illustrate the resulting point target and lesion images at 75 mm depth and without any aberrating layer in the propagation medium. A SOS of 1540 m/s was used to reconstruct these images. The point target (Fig. \ref{fig:CMTUSangle}(a)) has a LR of 1.78 mm and the lesion (Fig. \ref{fig:CMTUSangle}(f)) is visible with a contrast of -16.78 dB and CNR of 0.846. Note that, while the lesion is easily identified from the background, it is difficult to delineate its edges.

\subsubsection{CoMTUS discontinuous effective aperture}
Fig. \ref{fig:CMTUSangle} shows the simulated PSF and lesion images from the same non-aberrating medium and for increasing effective aperture and gap of the CoMTUS system. The control case (1-probe) is used for comparison (Fig. \ref{fig:CMTUSangle}(a,f)). It can be seen that, the PSF depends on the size of the effective aperture and the gap between the probes. As expected, the central lobe of the PSF reduces in width with increase in size of the effective aperture and so does the speckle size. However, while at extended apertures the width of the main lobe decreases, the amplitude of the side lobes increases with the corresponding gap in the aperture, affecting contrast as can be seen in the lesion images (Fig. \ref{fig:CMTUSangle}(g-j)). 

\begin{figure}[h]
\centering
\includegraphics[scale=1]{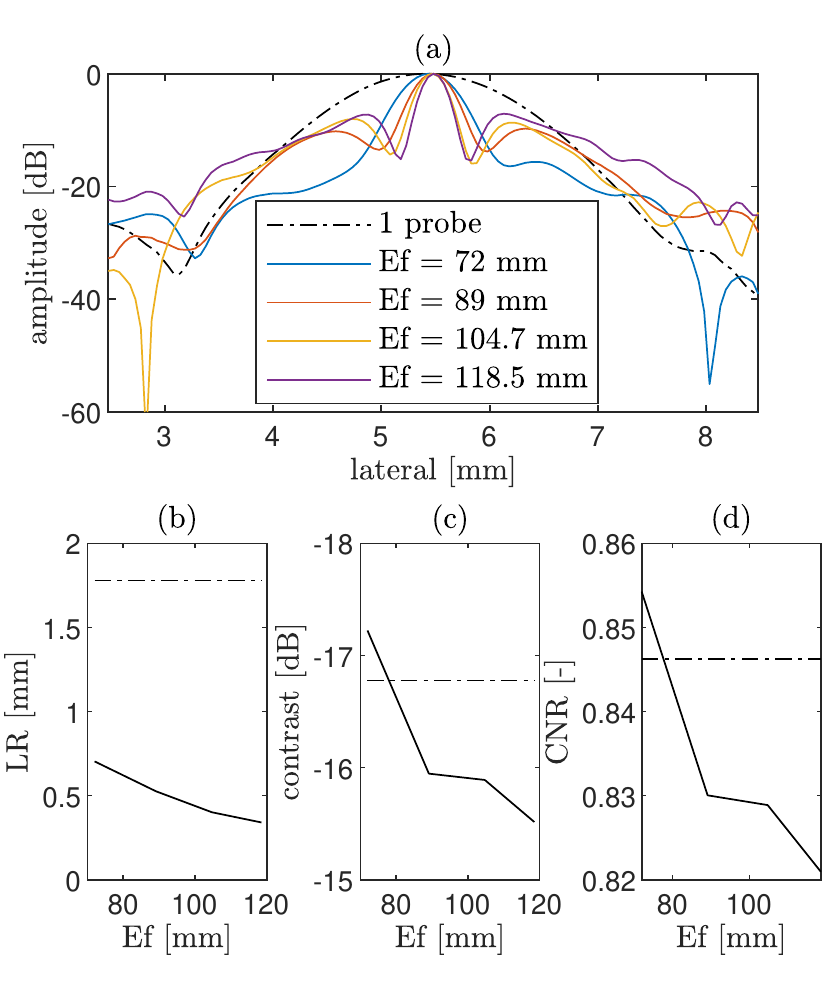}
\vspace{-0.5cm}
\caption{(a) Lateral PSF extracted from Fig. \ref{fig:CMTUSangle} at the depth of peak point intensity and in the principal direction. Corresponding computed quality metrics as function of the effective aperture size in CoMTUS (solid line) compared with 1-probe system (dashed line): (b) Lateral resolution (LR); (c) contrast and; (d) contrast-to-noise-ratio (CNR).} 
\label{fig:angle_metrics}
\end{figure}

The corresponding computed image quality metrics (LR, contrast and CNR) as function of the obtained effective aperture are compared with the 1-probe system in Fig. \ref{fig:angle_metrics}. Results show that both the main lobe of the PSF and the LR decrease with larger effective aperture size. Since an increasing effective aperture represents also a larger gap between the probes, contrast and resolution follow  opposite trends. In general, comparing with the 1-probe system, CoMTUS produces the best lateral resolution in all the cases but shows degradation in contrast with increasing probe separation.  
At the minimum separation CoMTUS produces an improvement in resolution compared to 1-probe of more than 50\% (0.70 mm vs 1.78 mm) combined with an improvement in contrast and CNR (-17.23 dB vs -16.78 dB, and 0.854 vs 0.846, respectively). At the maximum effective aperture simulated, resolution is the best with 0.34 mm, while the contrast and CNR drop to a minimum of -15.51 dB and 0.82, respectively.

\subsubsection{CoMTUS image penetration}
\begin{figure}
\centering
\includegraphics[width=1\linewidth]{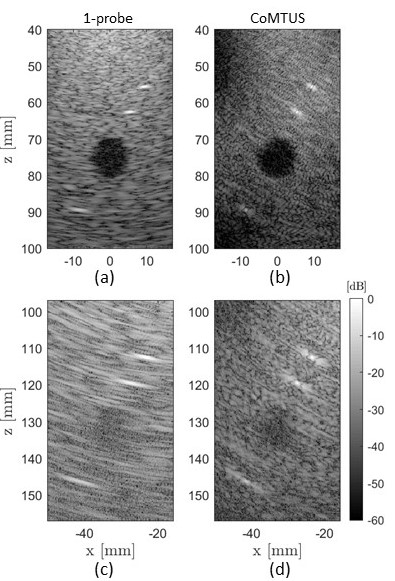}
\vspace{-0.5cm}
\caption{Comparison of simulated images acquired by a conventional aperture, 1-probe (a,c), and CoMTUS method (b,d) of the non-aberrating medium at different imaging depths.}
\label{fig:Depth}
\end{figure}

Fig. \ref{fig:Depth} compares the 1-probe with the CoMTUS (gap $\sim$45.3 mm) images at two different imaging depths (100 mm and 155 mm). 
Image degradation with depth is clearly observed in all the cases. However at larger depths the 1-probe shows a greater level of degradation. At the maximum imaging depth shown (155 mm), the point targets and the lesion can still be clearly identified in the CoMTUS image while they are barely discernible in the 1-probe image.

\begin{figure}
\centering
\includegraphics[scale=1]{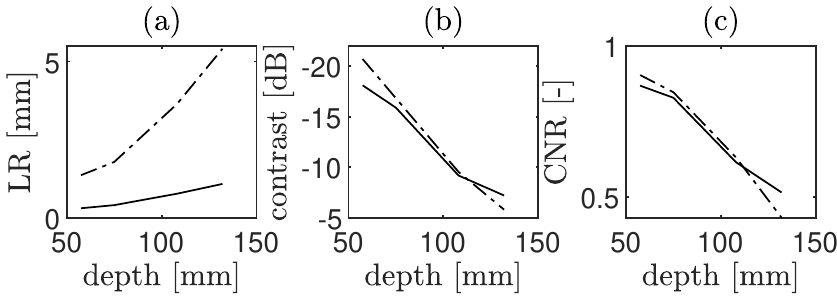}
\vspace{-0.2cm}
\caption{Computed quality metrics as function of the imaging depth: (a) Lateral resolution (LR); (b) contrast and; (c) contrast-to-noise-ratio (CNR). Two different methods are compared: 1-probe (dashed line) and CoMTUS (solid line).}
\label{fig:DepthQuanti}
\end{figure}

The computed image metrics as function of the imaging depth are summarized in Fig. \ref{fig:DepthQuanti}. As expected, in both systems, all image metrics worsen at larger imaging depths.  
Nevertheless, results show that their dependence on the imaging depth is different. LR is always larger (worse) with 1 probe and increases more rapidly with depth (Fig. \ref{fig:DepthQuanti}(a)). While at reduced imaging depths ($<$ 100 mm) contrast and CNR seem to be affected in a similar way in both systems, the loss in contrast metrics are less accentuated in the CoMTUS system at depths larger than 100 mm. In those cases, the CoMTUS method exceeds the performance of the 1-probe system not only in terms of resolution but also in contrast, despite in the used configuration the gap in the aperture is not minimum and leads to greater side lobes (Fig. \ref{fig:CMTUSangle}(d)).
The extended effective aperture created by CoMTUS consequently increases the sensitivity of the imaging system, particularly at large imaging depths.  

\subsubsection{CoMTUS through layered-media}
\begin{figure}
\centering
\includegraphics[scale=1]{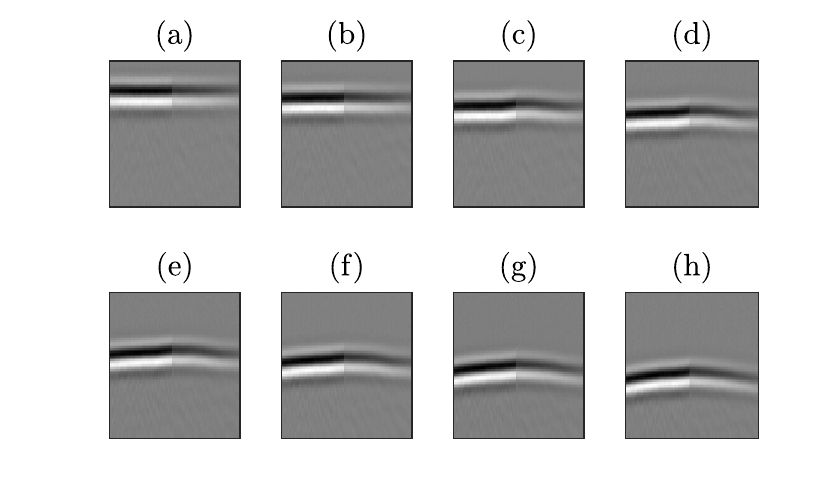}
\vspace{-0.5cm}
\caption{Simulated received wavefronts by the effective aperture after geometric compensation. Wavefronts are shown on a linear gray scale with time as the vertical axis and element number as the horizontal axis. The temporal range shown is 2.92 ms for 288 elements. Cases with different thickness of fat layer are shown: (a) 0 mm (control case); (b) 5 mm; (c) 10 mm; (d) 15 mm; (e) 20 mm; (f) 25 mm; (g) 30 mm; (h) 35 mm.}
\label{fig:aberration_wavefront}
\end{figure}

Aberration parameters measured from the simulated data are listed in Table \ref{tab:aberration properties}. ATF and ELF are defined relative to the arrival time and energy profiles expected for an undistorted spherical wavefront. Comparing with the range of rms ATF values from previous studies \cite{lacefield2002distributed}, the aberrating strength introduced by the layered-model can be considered from “weak” to “moderate”. Received wavefronts after geometric compensation are shown in Fig.\ref{fig:aberration_wavefront}. Shape distortions, resulted from propagation of the wavefront through the tissue layers, are apparent in the receiving aperture in all cases.
\begin{table}[ht]
\caption{Aberration parameters. Root-mean-squared (rms) values of and full-width at half-maximum correlation lengths (CL) of arrival time and energy level fluctuations (ATF and ELF, respectively).} 
\vspace{-0.5cm}
\begin{center}      
\begin{tabular}{|l|c|c|c|c|} 
\hline
\rule[-1ex]{0pt}{3.5ex} Fat+muscle & rms ATF & ATF CL  & rms ELF & ELF CL  \\
\rule[-1ex]{0pt}{3.5ex} thickness  [mm] & [ns] & [mm]  & [dB] & [mm]  \\
\hline
\rule[-1ex]{0pt}{3.5ex} 5+8 & 0.94 & 0.35 & 2.48 & 4.86 \\
\hline 
\rule[-1ex]{0pt}{3.5ex} 10+8 & 13.74  & 8.17 &  2.49 & 4.84 \\
\hline 
\rule[-1ex]{0pt}{3.5ex} 15+8 & 22.09  & 9.75  & 2.48  & 4.84\\
\hline 
\rule[-1ex]{0pt}{3.5ex} 20+8 & 23.03 & 8.76 & 2.47 & 4.82 \\
\hline 
\rule[-1ex]{0pt}{3.5ex} 25+8 & 32.53 & 9.34 & 2.47 & 4.80 \\
\hline 
\rule[-1ex]{0pt}{3.5ex} 30+8 & 35.64 & 7.44 & 2.48 & 4.80 \\
\hline 
\rule[-1ex]{0pt}{3.5ex} 35+8 & 39.64 & 9.37 & 2.49 & 4.75 \\
\hline 
\end{tabular}
\end{center}
\label{tab:aberration properties}
\end{table}

\begin{figure}
\centering
\includegraphics[width=1\linewidth]{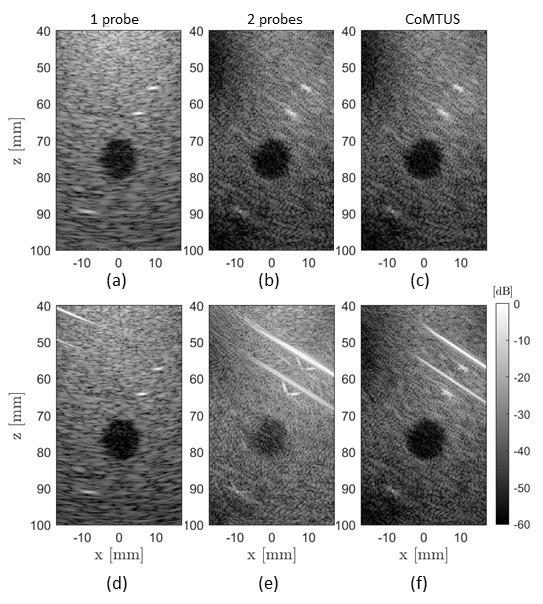}
\vspace{-0.5cm}
\caption{Comparison of simulated images for the control case (a,b,c) and imaging through fat and muscle layers of 35 mm and 8 mm thickness, respectively (d,e,f). Different methods are compared: 1-probe, 2-probes and CoMTUS.}
\label{fig:aberration_images}
\end{figure}

Fig. \ref{fig:aberration_images} shows the simulated images for the control case (medium only with soft tissue) and for imaging through fat and muscle layers of 35 mm and 8 mm thickness. respectively. 
The different methods (i.e. 1-probe, 2-probes coherently combined assuming probe locations are known and the SOS is constant, and fully optimised CoMTUS) are compared.  It can be seen that in the presence of aberration, the PSF and contrast of the 2-probes image significantly degrade when comparing with the control case.  
Indeed, in the presence of aberration, it is not possible to coherently reconstruct the image using the two separate transducers with fixed parameters (Fig. \ref{fig:aberration_images}(e)). However the CoMTUS system retains its performance (Fig. \ref{fig:aberration_images}(f)). 

\begin{figure}
\centering
\includegraphics[width=1\linewidth]{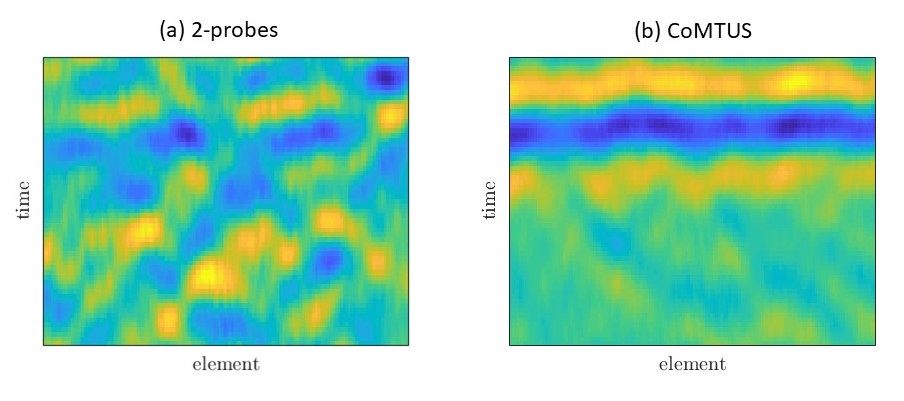}
\vspace{-0.8cm}
\caption{Simulated delayed RF data for a medium with a fat layer of 35 mm thickness and backscattered from a point-like target, obtained by coherently adding the 4 delayed backscattered echoes from the same point-like target ($T_1R_1,T_1R_2,T_2R_1,T_2R_2$) using different beamforming parameters: (a) 2-probes; (b) CoMTUS.}
\label{fig:delayedRFSimulation}
\end{figure}

Fig. \ref{fig:delayedRFSimulation} shows an example of the delayed echos from the point-like target for the 2-probes and CoMTUS cases, corresponding to a propagation medium with a fat layer of 35 mm thickness. These flat echoes are obtained by coherently adding the 4 delayed backscattered echoes from the same point-like target ($T_1R_1,T_1R_2,T_2R_1,T_2R_2$) using equation (\ref{eq:finalImage}) and the corresponding beamforming parameters. In the 2-probes case, the different echoes do not properly align, creating interference when forming a coherent addition of signals. However, after optimizing the beamforming parameters in the CoMTUS, all echos substantially align and can be coherently added together, minimizing the aberrating consequences. Similar effects are seen in the anechoic lesion. While differences in the background speckle pattern are observed between the different imaging methods, a higher loss of contrast due to aberration can be appreciated only in the 2-probes images. No significant changes in image quality resulting from aberration are appreciated in either the 1-probe or CoMTUS systems. Although both systems are able to image through aberrating layers, they show clear differences. The CoMTUS shows more detailed images than the 1-probe system. The speckle size is reduced and the different tissue layers are only visible in the CoMTUS images.  

\begin{figure}
\centering
\includegraphics[scale=1]{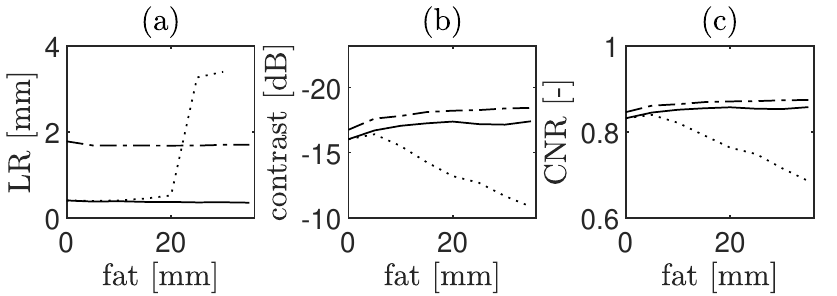}
\vspace{-0.2cm}
\caption{Computed quality metrics: (a) lateral resolution (LR); (b) contrast and; (c) contrast-to-noise-ratio (CNR), as function of the clutter thickness (fat layer). Three different methods are compared: 1-probe (solid line), 2-probes (dot line) and CoMTUS (dashed line).}
\label{fig:CMTUSaberration_metrics}
\end{figure}

The corresponding imaging metrics as function of fat layer thickness are shown in Fig. \ref{fig:CMTUSaberration_metrics}. As expected, in the absence of aberration, resolution improves with increasing aperture size. In this case, the worst lateral resolution corresponds to the 1-probe system with 1.78 mm, which is the one with smallest aperture size, while the 2-probes and CoMTUS images are similar with 0.40 mm.
The trends show that if aberration is not corrected, there are no significant improvements in the imaging metrics related to the aperture size for thicker thickness of fat layers. At clutter thickness larger than 10 mm, image quality of the system formed by 2 transducers without aberration correction (2-probes) is significantly degraded, while CoMTUS imaging metrics are not affected by aberration errors, following the same  trend as a conventional aperture (1-probe) and providing a constant value of resolution over clutter thickness without any significant loss of contrast. At the thickest fat layer simulated, resolution is 1.7 mm and 0.35 mm for the 1-probe and CoMTUS images, respectively,  while in the case of 2-probes images is no longer possible to reconstruct point-targets with sufficient definition to measure resolution. Contrast and CNR also show a similar significant loss for the 2-probes images. With the thickest aberrating layer, 2 probes case gives contrast of -10.84 dB and CNR of 0.69, while those values are significantly better for the 1-probe (-18.44 dB contrast and 0.87 CNR) and CoMTUS (-17.41 dB contrast and 0.86 CNR) images. 


\subsection{Experimental results}
\begin{figure}
\centering
\includegraphics[scale=1]{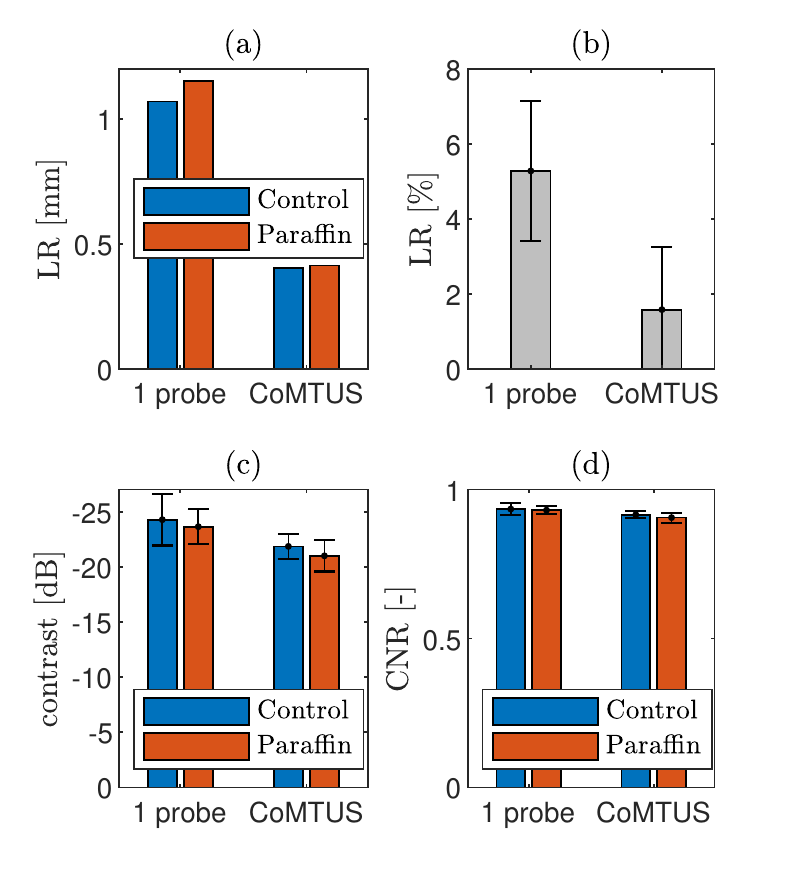}
\caption{Computed quality metrics experimentally measured. (a) Lateral resolution (LR) measured at 85 mm depth; (b) Changes in LR measured in the 4 point targets when imaging through the paraffin and relative to the control case; (c) contrast and; (d) contrast-to-noise-ratio (CNR). Two different methods are compared: 1-probe and CoMTUS.}
\label{fig:experiments_metrics}
\end{figure}

Coherent PW imaging with a conventional aperture (using a single probe) provides the reference for image quality with and without the paraffin wax layer. 
\begin{figure}
\centering
\includegraphics[width=\linewidth]{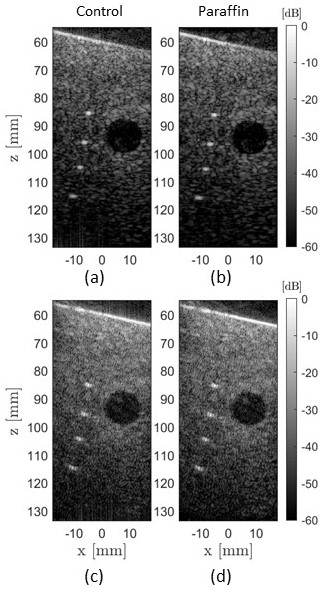}
\vspace{-0.5cm}
\caption{Experimental images of the control (a,c) and the paraffin cases (b,d). Two different methods are compared: 1-probe (a,b) and CoMTUS (c,d).}
\label{fig:experiments}
\end{figure}
To reconstruct these images the reference SOS in water of 1496 m/s was used and 7 PWs were compounded. Fig. \ref{fig:experiments} shows a comparison of the phantom images acquired with 1-probe and CoMTUS in the control case and through the paraffin wax sample. The CoMTUS images were reconstructed using the adaptively optimised beamfoming parameters, which include the average SOS and compounding 6 PWs. All images are shown in the same dynamic range of -60 dB. In both cases, 1-probe and CoMTUS images, little variation is observed between the control and the paraffin images, which is consistent with the simulation results.  
The values of the optimum beamforming parameters used to reconstruct the CoMTUS images are $\{c=1488.5 \;m/s, \theta_2 = 30.04^{o}, \mathbf{r}_2 = [46.60, 12.33] \;mm \}$ for the control case and  $\{c = 1482.6 \;m/s, \theta_2 = 30.00^{o}, \mathbf{r}_2 = [46.70, 12.37] \;mm \}$ for the paraffin. There are slight changes in all the values and a drop in the average SOS which agrees with the lower SOS of the paraffin wax. 
Fig. \ref{fig:experiments_metrics} summarizes the computed image metrics for both the control and the paraffin cases. 
Little variation was observed in all the imaging metrics. Although minimum image degradation by aberrating layers was observed in the CoMTUS, the overall image quality (particularly resolution) improved compared with the conventional single aperture and the observed image degradation follows the same trend.

\begin{figure}
\centering
\includegraphics[width=\linewidth]{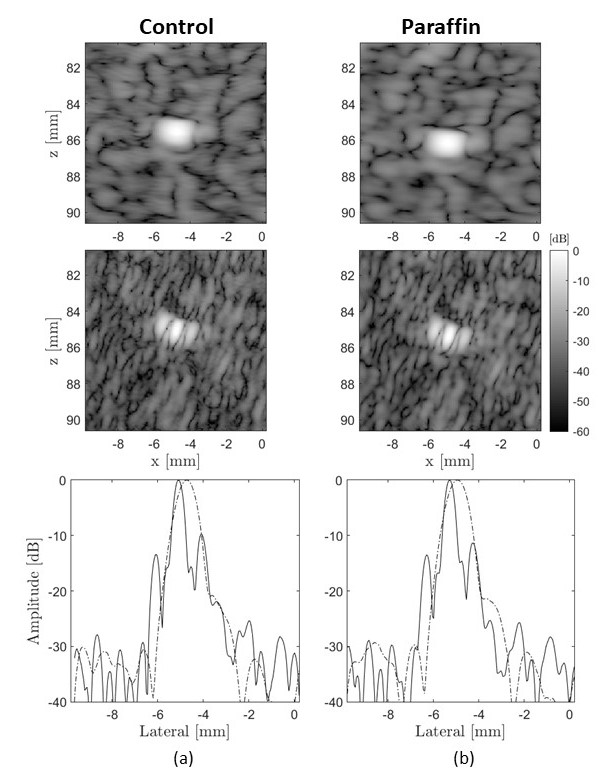}
\caption{Experimental point target images. Column (a) corresponds to the control and column (b) to the paraffin. First row corresponds to 1-probe system and middle row to CoMTUS. Bottom row shows the corresponding lateral point spread functions for the two cases displayed: 1-probe system (dashed line) and CoMTUS (solid line).}
\label{fig:PSFexperiments}
\end{figure}

\begin{figure}
\centering
\includegraphics[width=\linewidth]{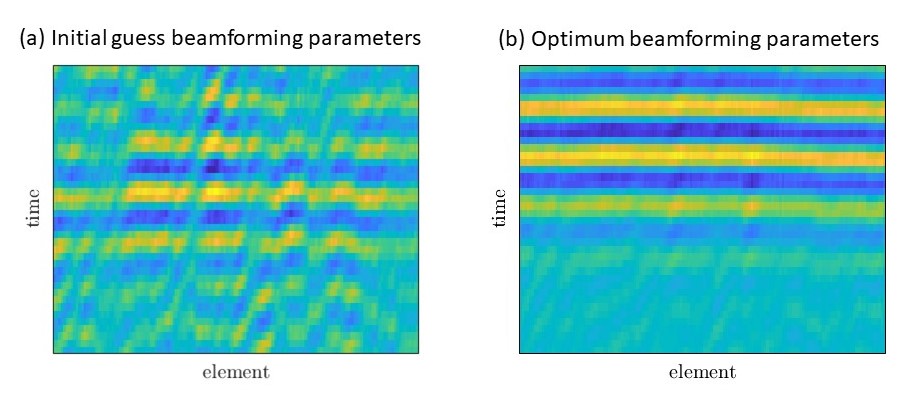}
\caption{Experimental delayed RF data acquired from the phantom with the paraffin wax sample. CoMTUS flat backscattered echo from a point-like target, obtained by coherently adding the 4 delayed backscattered echoes from the same point-like target ($T_1R_1,T_1R_2,T_2R_1,T_2R_2$) using different beamforming parameters: (a) initial guess values; (b) optimum values.}
\label{fig:delayedRF}
\end{figure}

The first point target located at 85 mm depth was described using its lateral PSF (Fig. \ref{fig:PSFexperiments}), with and without the paraffin wax layer. No significant effects due to aberration are observed in the PSF in any of the cases. The PSF shape is similar with and without the paraffin wax layer and agrees with the one observed in simulations. In general the CoMTUS method leads to a PSF with significant narrower main lobe but also with side lobes of bigger amplitude than the 1-probe conventional imaging system. 
Fig. \ref{fig:delayedRF} shows the coherent summation of the delayed echos from the point-like target before and after optimization. The effects of the paraffin layer are clearly seen. When the beamforming parameters, including the averaged SOS, are optimized by the CoMTUS method, all echos align better, minimizing the aberrating paraffin effects. 

\section{\label{sec:5}DISCUSSION}
The implications for imaging using the CoMTUS method with two linear arrays have been further investigated here with simulations and experiments. This analysis has shown that, the performance of CoMTUS depends on the relative location of the sub-arrays, the CoMTUS sensitivity increases with the imaging depth and the resulting extended aperture preserves in layered-media with different SOS. These findings show that, if the separation between the transducers is limited, the extended effective aperture created by CoMTUS confers benefits in resolution and contrast that improve image quality, particularly at large imaging depths, compared with a single probe standard system, and even in the presence of acoustic clutter imposed by tissue layers of different SOS (Fig. \ref{fig:aberration_images}, \ref{fig:experiments}). Extra analyses would be needed to quantify the significance of the results in different scenarios and further support these claims. 

Unlike the improvement achieved in resolution, benefits in contrast are not so significant. Indeed simulation results suggest that, the discontinuous effective aperture may degrade contrast when the gap in the aperture is bigger than a few centimeters. In probe design, there is a requirement of half wavelength spacing between elements in order to avoid the occurrence of unwanted grating lobes in the array response \cite{lee1988antenna}. 
Moreover, previous studies indicated that, unlike resolution, contrast does not continue to increase uniformly at larger aperture sizes \cite{bottenus2016large,bottenus2018evaluation}. Nevertheless, while the contrast may be degraded by big discontinuities in the aperture, the main lobe resolution continues to improve at larger effective apertures. Since the lesion detectability is a function of both the contrast and resolution \cite{karaman1995synthetic}, overall there are benefits from extended aperture size, even when contrast is limited. A narrow main lobe allows fine sampling of high resolution targets, providing improved visibility of edges of clinically relevant targets. Both numerical and experimental results show how the speckle size significantly reduces when there is an enlarged receiving aperture produced by coherent combination of multiple probes (Fig. \ref{fig:CMTUSangle} and \ref{fig:PSFexperiments}). In addition, when imaging at larger depths, an extended aperture has the potential to improve the attenuation-limited image quality. In those challenging cases at large imaging depths, CoMTUS shows improvements not only in resolution but also in contrast, even for effective apertures with significant gaps that rise the sidelobes (Fig. \ref{fig:Depth} and \ref{fig:DepthQuanti}).

Our results agree with the hypothesis that in the absence of aberration, the aperture size determines resolution \cite{cobbold2006foundations}.  However, previous works suggest that despite predicted gains in resolution, there are practical limitations to the gains made at larger aperture sizes \cite{moshfeghi1988vivo}. Inhomogeneities caused changes in the sidelobes and focal distance, limiting the improvement in resolution. The resulting degradation is primarily due to the arrival time variation called phase aberration.  The outer elements on a large transducer suffer from severe phase errors due to an aberrating layer of varying thickness, placing limits on the gains to be made from large arrays \cite{trahey1991vivo}. Findings presented here agree with these previous studies when there are fixed average parameters (coherent 2-probes case) in the presence of aberration clutter. However, the adaptive optimisation deployed in the CoMTUS method takes into account the average SOS along the propagation path in the medium and shows promise for extending the effective aperture beyond this conventional practical limit imposed by the clutter. More accurate SOS estimation improves beamforming and allows higher order phase aberration correction.
However others challenges imposed by aberration still remain. Recent studies reveal that both phase aberration and reverberation are primary contributors to degraded image quality \cite{pinton2014spatial}. While phase aberration effects are caused by variations in SOS due to tissue inhomogeneity, reverberation is caused by multiple reflections within inhomogeneous medium, generating clutter that distorts the appearance of the wavefronts from the region of interest. For fundamental imaging, reverberations have been shown to be a significant cause of image quality degradation and are the principal reason why harmonic US imaging is better than fundamental imaging \cite{fatemi2019studying}. Since CoMTUS relies on detecting backscattered echoes from targeted scattered points, the optimization may be corrupted if the reverberation clutter overlays these echoes. The magnitude of this effect will depend mostly on the clutter strength and the signal-to-noise ratio. For all the cases described here, it was possible to detect the backscattered echoes of the targets used for calibration and CoMTUS optimization succeeded. 
However, there may be more complex situations in vivo and further investigations are needed.
Future work will focus on the role of redundancy in the large array in averaging multiple realizations of the reverberation signal as a mechanism for clutter reduction. 

Finally, it should be acknowledged that some choices made in the design of this study may not directly translate to clinical practice, but we believe that they do not compromise the conclusions drawn from these results. The available experimental setup drove the election of the frequency, which was higher than is traditionally used in abdominal imaging (1-2 MHz). In addition, both the simulated and experimental phantoms are a rather simplistic model of real human tissue.
Simulations were limited to a 2-dimensional model while real ultrasound propagation occurs in 3 dimensions. In 3 dimensions, extra clutter can be generated also by off-axis scattering, arising from scatterers located away from the beam's axis. Second, although tissue properties were taken from literature, it is well known that ultrasonic tissue properties vary considerably among individual tissue specimens and different measurement techniques. Finally, the main limitation is the lack of tissue microstructure in the model. Only with a more complete description of tissue microstructure it is possible to capture the phase and amplitude errors observed in vivo in pulse echo measurements. However, results presented here can capture the main time-shift fluctuations and then help in further understanding of CoMTUS. Previous simulation studies have shown that a simple model like the one used in this study successfully predicts the magnitudes and large-scale trends of time-shift fluctuations, but not the energy-level fluctuations and waveform distortions \cite{mast1997simulation}. Further studies using more sophisticated models are needed to describe the wave distortions, including phase and energy fluctuations.

\section{\label{sec:6}Conclusion}
In this study, the implications for imaging using the CoMTUS method have been further investigated by simulation and experiments.  
A CoMTUS system formed by two identical linear arrays has been simulated using the k-wave Matlab toolbox to analyze the implications of different spatial parameters for imaging. A parametric study is presented, where the resulting effective aperture size, separation between the arrays, imaging depth and the presence of acoustic clutter have been investigated.   
Results show that, the performance of CoMTUS depends on the relative location of the arrays, that CoMTUS sensitivity increases with imaging depth and that the resulting extended aperture maintains its performance when imaging through a layered-medium with different speed of sound. Both simulated and experimental results show that CoMTUS improves US imaging quality in terms of resolution and shows higher sensitivity with imaging depth, providing benefits to image quality compared with conventional 1-probe imaging systems.


%



\section*{Acknowledgment}
This work was supported by the Wellcome Trust/EPSRC iFIND project, IEH Award [102431] (www.iFINDproject.com) and the Wellcome Trust/EPSRC funded Centre for Medical Engineering [WT 203148/Z/16/Z]. The authors acknowledge financial support from the Department of Health via the National Institute for Health Research (NIHR) comprehensive
Biomedical Research Centre award to Guy's \& St Thomas' NHS Foundation Trust in partnership with King's College London and Kings College Hospital NHS Foundation Trust.

\ifCLASSOPTIONcaptionsoff
  \newpage
\fi


\begin{thebibliography}{10}
\providecommand{\url}[1]{#1}
\csname url@samestyle\endcsname
\providecommand{\newblock}{\relax}
\providecommand{\bibinfo}[2]{#2}
\providecommand{\BIBentrySTDinterwordspacing}{\spaceskip=0pt\relax}
\providecommand{\BIBentryALTinterwordstretchfactor}{4}
\providecommand{\BIBentryALTinterwordspacing}{\spaceskip=\fontdimen2\font plus
\BIBentryALTinterwordstretchfactor\fontdimen3\font minus
  \fontdimen4\font\relax}
\providecommand{\BIBforeignlanguage}[2]{{%
\expandafter\ifx\csname l@#1\endcsname\relax
\typeout{** WARNING: IEEEtran.bst: No hyphenation pattern has been}%
\typeout{** loaded for the language `#1'. Using the pattern for}%
\typeout{** the default language instead.}%
\else
\language=\csname l@#1\endcsname
\fi
#2}}
\providecommand{\BIBdecl}{\relax}
\BIBdecl

\bibitem{cobbold2006foundations}
R.~S. Cobbold, \emph{Foundations of biomedical ultrasound}.\hskip 1em plus
  0.5em minus 0.4em\relax Oxford University Press, 2006.

\bibitem{harris1991ultimate}
R.~A. Harris, D.~Follett, M.~Halliwell, and P.~Wells, ``Ultimate limits in
  ultrasonic imaging resolution,'' \emph{Ultrasound in medicine \& biology},
  vol.~17, no.~6, pp. 547--558, 1991.

\bibitem{moshfeghi1988vivo}
M.~Moshfeghi and R.~Waag, ``In vivo and in vitro ultrasound beam distortion
  measurements of a large aperture and a conventional aperture focussed
  transducer,'' \emph{Ultrasound in Medicine and Biology}, vol.~14, no.~5, pp.
  415--428, 1988.

\bibitem{tsai2015obesity}
P.-J.~S. Tsai, M.~Loichinger, and I.~Zalud, ``Obesity and the challenges of
  ultrasound fetal abnormality diagnosis,'' \emph{Best Practice \& Research
  Clinical Obstetrics \& Gynaecology}, vol.~29, no.~3, pp. 320--327, 2015.

\bibitem{klysik2014challenges}
M.~Klysik, S.~Garg, S.~Pokharel, J.~Meier, N.~Patel, and K.~Garg, ``Challenges
  of imaging for cancer in patients with diabetes and obesity,'' \emph{Diabetes
  technology \& therapeutics}, vol.~16, no.~4, pp. 266--274, 2014.

\bibitem{bottenus2016feasibility}
N.~Bottenus, W.~Long, H.~K. Zhang, M.~Jakovljevic, D.~P. Bradway, E.~M. Boctor,
  and G.~E. Trahey, ``Feasibility of swept synthetic aperture ultrasound
  imaging,'' \emph{IEEE transactions on medical imaging}, vol.~35, no.~7, pp.
  1676--1685, 2016.

\bibitem{bottenus2018evaluation}
N.~Bottenus, W.~Long, M.~Morgan, and G.~Trahey, ``Evaluation of large-aperture
  imaging through the ex vivo human abdominal wall,'' \emph{Ultrasound in
  medicine \& biology}, vol.~44, no.~3, pp. 687--701, 2018.

\bibitem{zimmer2018}
V.~A. Zimmer, A.~Gomez, Y.~Noh, N.~Toussaint, B.~Khanal, R.~Wright, L.~Peralta,
  M.~van Poppel, E.~Skelton, J.~Matthew, and J.~A. Schnabel, ``Multi-view image
  reconstruction: Application to fetal ultrasound compounding,'' in \emph{Data
  Driven Treatment Response Assessment and Preterm, Perinatal, and Paediatric
  Image Analysis}, A.~Melbourne, R.~Licandro, M.~DiFranco, P.~Rota, M.~Gau,
  M.~Kampel, R.~Aughwane, P.~Moeskops, E.~Schwartz, E.~Robinson, and
  A.~Makropoulos, Eds.\hskip 1em plus 0.5em minus 0.4em\relax Cham: Springer
  International Publishing, 2018, pp. 107--116.

\bibitem{peralta2018ius}
L.~Peralta, A.~Gomez, J.~V. Hajnal, and R.~J. Eckersley, ``Feasibility study of
  a coherent multi-transducer us imaging system,'' in \emph{2018 IEEE
  International Ultrasonics Symposium (IUS)}.\hskip 1em plus 0.5em minus
  0.4em\relax IEEE, 2018, pp. 1--4.

\bibitem{peralta2019}
L.~Peralta, A.~Gomez, J.~V. Hajnal, Y.~Luan, B.~Kim, and R.~J. Eckersley,
  ``Coherent multi-transducer ultrasound imaging,'' \emph{IEEE Transactions on
  Ultrasonics, Ferroelectrics and Frequency Control}, vol.~66, no.~8, pp.
  1316--1330, 2019.

\bibitem{peralta2019SPIE}
L.~Peralta, A.~Gomez, J.~V. Hajnal, and R.~J. Eckersley, ``Coherent
  multi-transducer ultrasound imaging in the presence of aberration,'' in
  \emph{Medical Imaging 2019: Ultrasonic Imaging and Tomography}, vol.
  10955.\hskip 1em plus 0.5em minus 0.4em\relax International Society for
  Optics and Photonics, 2019, p. 109550O.

\bibitem{peralta2019iusaberration}
L.~Peralta, M.~Reinwald, J.~V. Hajnal, and R.~J. Eckersley, ``Coherent
  multi-transducer ultrasound imaging through aberrating media,'' in \emph{2019
  IEEE International Ultrasonics Symposium (IUS)}.\hskip 1em plus 0.5em minus
  0.4em\relax IEEE, 2019, pp. 324--327.

\bibitem{peralta2019iusextension}
L.~Peralta, M.~Reinwald, A.~Ramalli, J.~V. Hajnal, and R.~J. Eckersley,
  ``Extension of coherent multi-transducer ultrasound imaging with diverging
  waves,'' in \emph{2019 IEEE International Ultrasonics Symposium (IUS)}.\hskip
  1em plus 0.5em minus 0.4em\relax IEEE, 2019, pp. 2226--2229.

\bibitem{fitzgibbon2003robust}
A.~W. Fitzgibbon, ``Robust registration of {2D} and {3D} point sets,''
  \emph{Image and Vision Computing}, vol.~21, no. 13-14, pp. 1145--1153, 2003.

\bibitem{Montaldo2009CoherentElastography}
\BIBentryALTinterwordspacing
G.~Montaldo, M.~Tanter, J.~Bercoff, N.~Benech, and M.~Fink, ``{Coherent
  plane-wave compounding for very high frame rate ultrasonography and transient
  elastography},'' \emph{IEEE Transactions on Ultrasonics, Ferroelectrics and
  Frequency Control}, vol.~56, no.~3, pp. 489--506, 3 2009. [Online].
  Available: \url{http://ieeexplore.ieee.org/document/4816058/}
\BIBentrySTDinterwordspacing

\bibitem{christensen2014vivo}
K.~Christensen-Jeffries, R.~J. Browning, M.-X. Tang, C.~Dunsby, and R.~J.
  Eckersley, ``In vivo acoustic super-resolution and super-resolved velocity
  mapping using microbubbles,'' \emph{IEEE transactions on medical imaging},
  vol.~34, no.~2, pp. 433--440, 2014.

\bibitem{christensen2019coherent}
K.~Christensen-Jeffries, L.~Peralta, M.~Reinwald, J.~V. Hajnal, and R.~J.
  Eckersley, ``Coherent multi-transducer ultrasound imaging with micro-bubble
  contrast agents,'' in \emph{2019 IEEE International Ultrasonics Symposium
  (IUS)}.\hskip 1em plus 0.5em minus 0.4em\relax IEEE, 2019, pp. 2310--2312.

\bibitem{treeby2012modeling}
B.~E. Treeby, J.~Jaros, A.~P. Rendell, and B.~Cox, ``Modeling nonlinear
  ultrasound propagation in heterogeneous media with power law absorption using
  a k-space pseudospectral method,'' \emph{The Journal of the Acoustical
  Society of America}, vol. 131, no.~6, pp. 4324--4336, 2012.

\bibitem{treeby2010k}
B.~E. Treeby and B.~T. Cox, ``{k-Wave}: Matlab toolbox for the simulation and
  reconstruction of photoacoustic wave fields,'' \emph{Journal of biomedical
  optics}, vol.~15, no.~2, p. 021314, 2010.

\bibitem{denarie2013coherent}
B.~Denarie, T.~A. Tangen, I.~K. Ekroll, N.~Rolim, H.~Torp, T.~Bj{\aa}stad, and
  L.~Lovstakken, ``Coherent plane wave compounding for very high frame rate
  ultrasonography of rapidly moving targets,'' \emph{IEEE Transactions on
  Medical Imaging}, vol.~32, no.~7, pp. 1265--1276, 2013.

\bibitem{Brown2019Detection}
J.~Brown, K.~Christensen-Jeffries, S.~Harput, G.~Zhang, J.~Zhu, C.~Dunsby,
  M.~Tang, and R.~Eckersley, ``Investigation of microbubble detection methods
  for super-resolution imaging of microvasculature,'' \emph{IEEE Transactions
  on Ultrasonics, Ferroelectrics and Frequency Control}, 2019.

\bibitem{treeby2010a}
B.~E. Treeby and B.~Cox, ``Modeling power law absorption and dispersion for
  acoustic propagation using the fractional laplacian,'' \emph{The Journal of
  the Acoustical Society of America}, vol. 127, no.~5, pp. 2741--2748, 2010.

\bibitem{pinton2009heterogeneous}
G.~F. Pinton, J.~Dahl, S.~Rosenzweig, and G.~E. Trahey, ``A heterogeneous
  nonlinear attenuating full-wave model of ultrasound,'' \emph{IEEE
  transactions on ultrasonics, ferroelectrics, and frequency control}, vol.~56,
  no.~3, 2009.

\bibitem{robertson2017accurate}
J.~L. Robertson, B.~T. Cox, J.~Jaros, and B.~E. Treeby, ``Accurate simulation
  of transcranial ultrasound propagation for ultrasonic neuromodulation and
  stimulation,'' \emph{The Journal of the Acoustical Society of America}, vol.
  141, no.~3, pp. 1726--1738, 2017.

\bibitem{wise2017staircase}
E.~S. Wise, J.~L. Robertson, B.~T. Cox, and B.~E. Treeby, ``Staircase-free
  acoustic sources for grid-based models of wave propagation,'' in \emph{2017
  IEEE International Ultrasonics Symposium (IUS)}.\hskip 1em plus 0.5em minus
  0.4em\relax IEEE, 2017, pp. 1--4.

\bibitem{goss1978comprehensive}
S.~Goss, R.~Johnston, and F.~Dunn, ``Comprehensive compilation of empirical
  ultrasonic properties of mammalian tissues,'' \emph{The Journal of the
  Acoustical Society of America}, vol.~64, no.~2, pp. 423--457, 1978.

\bibitem{liu1994correction}
D.-L. Liu and R.~C. Waag, ``Correction of ultrasonic wavefront distortion using
  backpropagation and a reference waveform method for time-shift
  compensation,'' \emph{The Journal of the Acoustical Society of America},
  vol.~96, no.~2, pp. 649--660, 1994.

\bibitem{mast1997simulation}
T.~D. Mast, L.~M. Hinkelman, M.~J. Orr, V.~W. Sparrow, and R.~C. Waag,
  ``Simulation of ultrasonic pulse propagation through the abdominal wall,''
  \emph{The Journal of the Acoustical Society of America}, vol. 102, no.~2, pp.
  1177--1190, 1997.

\bibitem{boni2016ula}
E.~Boni, L.~Bassi, A.~Dallai, F.~Guidi, V.~Meacci, A.~Ramalli, S.~Ricci, and
  P.~Tortoli, ``{ULA-OP} 256: A 256-channel open scanner for development and
  real-time implementation of new ultrasound methods,'' \emph{IEEE Transactions
  on Ultrasonics, Ferroelectrics and Frequency Control}, vol.~63, no.~10, pp.
  1488--1495, 2016.

\bibitem{boni2017architecture}
E.~Boni, L.~Bassi, A.~Dallai, V.~Meacci, A.~Ramalli, M.~Scaringella, F.~Guidi,
  S.~Ricci, and P.~Tortoli, ``Architecture of an ultrasound system for
  continuous real-time high frame rate imaging,'' \emph{IEEE transactions on
  ultrasonics, ferroelectrics, and frequency control}, vol.~64, no.~9, pp.
  1276--1284, 2017.

\bibitem{smith1985frequency}
S.~Smith, H.~Lopez, and W.~Bodine~Jr, ``Frequency independent ultrasound
  contrast-detail analysis,'' \emph{Ultrasound in medicine \& biology},
  vol.~11, no.~3, pp. 467--477, 1985.

\bibitem{lacefield2002distributed}
J.~C. Lacefield, W.~C. Pilkington, and R.~C. Waag, ``Distributed aberrators for
  emulation of ultrasonic pulse distortion by abdominal wall,'' \emph{Acoustics
  Research Letters Online}, vol.~3, no.~2, pp. 47--52, 2002.

\bibitem{lee1988antenna}
S.~Lee and Y.~Lo, \emph{Antenna Handbook: theory, applications, and
  design}.\hskip 1em plus 0.5em minus 0.4em\relax Van Nostrand Reinhold, 1988.

\bibitem{bottenus2016large}
N.~Bottenus, G.~Pinton, and G.~Trahey, ``Large coherent apertures: Improvements
  in deep abdominal imaging and fundamental limits imposed by clutter,'' in
  \emph{2016 IEEE International Ultrasonics Symposium (IUS)}.\hskip 1em plus
  0.5em minus 0.4em\relax IEEE, 2016, pp. 1--4.

\bibitem{karaman1995synthetic}
M.~Karaman, P.-C. Li, and M.~O'Donnell, ``Synthetic aperture imaging for small
  scale systems,'' \emph{IEEE transactions on ultrasonics, ferroelectrics, and
  frequency control}, vol.~42, no.~3, pp. 429--442, 1995.

\bibitem{trahey1991vivo}
G.~Trahey, P.~Freiburger, L.~Nock, and D.~Sullivan, ``In vivo measurements of
  ultrasonic beam distortion in the breast,'' \emph{Ultrasonic imaging},
  vol.~13, no.~1, pp. 71--90, 1991.

\bibitem{pinton2014spatial}
G.~F. Pinton, G.~E. Trahey, and J.~J. Dahl, ``Spatial coherence in human
  tissue: Implications for imaging and measurement,'' \emph{IEEE transactions
  on ultrasonics, ferroelectrics, and frequency control}, vol.~61, no.~12, pp.
  1976--1987, 2014.

\bibitem{fatemi2019studying}
A.~Fatemi, E.~A.~R. Berg, and A.~Rodriguez-Molares, ``Studying the origin of
  reverberation clutter in echocardiography: In vitro experiments and in vivo
  demonstrations,'' \emph{Ultrasound in medicine \& biology}, 2019.

\end{thebibliography}
\end{document}